\documentclass[aps,prl,twocolumn,showpacs,a4paper,unsortedaddress,amsmath,amssymb,byrevtex]{revtex4}

\usepackage[latin1]{inputenc}
\usepackage{graphicx}
\usepackage{color}
\usepackage{dcolumn}
\usepackage{bm}
\usepackage{hyperref}
\usepackage{times}

\begin{document}

\title{Unconventional magnetism in small gold organic molecules}

\author{Diego Carrascal}
\author{Lucas Fern\'andez-Seivane}
\author{Jaime Ferrer}
\affiliation{
Departamento de Fisica, Universidad de Oviedo, 33007 Oviedo, Spain and
\\Centro de Investigaci\'on en Nanomateriales y Nanotecnolog\'{\i}a, CSIC,
Spain}

\date{\today}

\begin{abstract}
We present a theoretical study of the
magnetic properties of dicyclopentadienyl metallocene and phthalocyanine molecules,
that contain the transition metal atoms M = Fe, Co, Ni, Cu, Zn, Ir, Pt and Au. Our most important
prediction is that gold and copper molecules are magnetic. We find that the magnetism of these
molecules is fairly unconventional: the gold atom itself is weakly magnetic or even non-magnetic.
Its role is rather to induce magnetism in the surrounding carbon and nitrogen atoms, producing a sort
of spin density wave. 
\end{abstract}


\maketitle

The efforts devoted in the past few years to fabricate and characterize new nano-scale objects have
helped to uncover a wealth of fascinating geometric, mechanical, electronic, magnetic, optical
or dissipative properties, that are brought about in many cases by the laws of quantum physics. 
More specifically, the multidisciplinary field of Nanomagnetism aims at the fabrication of devices
with tailored magnetic properties.
Indeen, there exists an intense activity in the micromagnetic community to extend its reach to the
nanoscale,
by reducing the size of their room-temperature magnetic dots well below the 100 nm mark.
Gold nanoparticles of diameters in the 2 nanometer range have been found recently to show
room-temperature magnetism, when capped with organic 
molecules\cite{Crespo04,Yamamoto04,Garitaonandia08}. Interestingly, the experiments have
found that the magnetism of these nanoparticles is localized at their surface, where
Au atoms are in chemical contact with the capping molecules. The different
experiments produce different estimates for the spin moments $M_S$ of the surface  gold
atoms, which vary 
from 0.002 to 0.3 $\mu_B$\cite{Garitaonandia08}, but they agree on the very high value of the 
magnetic anisotropy energy (MAE), which is of about 0.4 eV/atom. Huge
orbital moments $M_L$ at the surface gold atoms have been proposed to exist and
originate that large MAE\cite{Hernando06}, but the actual measured moments seem to be modest,
of order $M_L/M_S \sim 0.15$\cite{Garitaonandia08,Yamamoto04}.

Molecular magnets are in some ways superior to nanoparticles, since they have a 
well defined number of atoms with precise chemical identities, and do not suffer from
particle number dispersion. Further, molecules are not so prone to conformational changes, since 
these require breaking a sizable number of covalent bonds, as opposed to the metallic bonds
of atomic clusters. Finally, the chemical activity of molecules can be engineered by
oxidation or reduction, or by the attachment  of end groups.
Molecular magnets containing single Rare-Earth ions have been studied in the 
past\cite{Gatteschi,Benelli02}, but  late 5d elements like iridium, platinum or gold are possibly
better candidates for room-temperature magnetism since these are heavier elements, which means that
they have higher Spin-Orbit coupling constants. Our previous calculations for platinum dimers found
MAEs of the order of 0.1 eV/atom\cite{Seivane07}, which are consistent with the experimental
values in gold nanoparticles referred above. 

We present here a detailed theoretical study of the magnetism of two popular organic molecules,
metallocenes and phthalocyanines. 
Metallocenes, denoted MCp$_2$, are organometallic molecules that contain two cyclopentadyenil rings
Cp, which sandwich a transition metal atom M. Metallic phthalocyanines, denoted MPc, are macrocyclic
molecules that have an alternating nitrogen atom-carbon atom ring structure, and contain a metal ion 
M in its center, which bonds with the four isoindole nitrogen atoms.
We have used the third row elements Fe, Co, Ni, Cu and Zn, and the fifth row elements Ir, Pt and Au
as the metallic atom M. 
The axial symmetry of these molecules produces a crystal field at the metal ion position which
leads to expect that Fe and Zn molecules have a net zero spin, that Co, Ir, Cu and Au molecules
have spin 1/2, and that Co and Pt molecules have spin 1. Our
simulations confirm these guesses  except for Pt molecules, that have a spinless ground
state. Interestingly, we find that gold molecules have a finite spin 1/2, but that the
distribution of magnetic moments across gold
molecules is not conventional. The gold atoms produce a sort of spin density wave such that 
the surrounding carbon or nitrogen atoms are also magnetized, showing a spin-density-wave
pattern, with almost collinear orientations. The gold atom itself contributes about 15 \% of the total
spin moment of the molecule in orocene, while it is unmagnetized in the phthalocyanine. The
orbital moments are between three and thirty times smaller that the spin moments. We have checked
that all the spin
1/2 molecules show zero magnetic anisotropy as they should\cite{c-number}. The spin-1 Ni molecules
also show very small anisotropies, which was expected because of its small Spin Orbit coupling
constant.

The structural and electronic properties of metallocene and possibly also of
phthalocyanine molecules in the gas and crystalline phases are well captured by density functional
theory (DFT)\cite{Swart07}. 
We have therefore carried the simulations with the molecular dynamics SIESTA
suite\cite{SIESTA}, which uses norm-conserving pseudopotentials and a basis set of localized
atomic-like wave functions.
We have used a triple-zeta doubly polarized basis set for transition metal ions (Fe, Co, Ni, Cu,
Zn, Ir, Pt and Au), and a double-zeta polarized basis set for the carbon, nitrogen and hydrogen
atoms. We have checked explicitly that our simulation boxes for metallocenes, having 20 $\AA$ of
lateral size, where large enough to avoid spurious boundary effects of the electric field. For 
the larger phthalocyanine molecules, the boxes were cuboids of size 28x28x20 \AA$^3$. We have used a
very fine grid for the real space integrals
to ensure that egg-box effects were absent. We found it essential to set a rather strict tolerance
of $5 \times 10^{-4}$ eV/\AA~ in the force relaxation procedure of the atoms, in order to scape
from local minima of the energy landscape. We took between 5 and 10 different initial seeds for 
the geometry/spin arrangements in our sampling of the Hilbert space of each quantum system to ensure
that we would reach the ground state configuration. We have confirmed that the Local Density and
the Generalized Gradient approximations (LDA and GGA, respectively)\cite{Per81,Per96} shed the same
results for the ground state and first excited isomers of all the molecules, with only small
quantitative differences.
We have also double-checked our results for gold molecules 
with the alternative code QUANTUM ESPRESSO\cite{PWSCF}, which uses ultra-soft pseudopotentials
and a plane-wave basis set.

\begin{figure}
\includegraphics[width=\columnwidth]{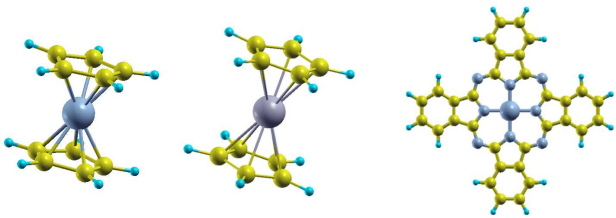}
\caption{(Color online) (a) eclipsed, and (b) staggered, geometries of metallocene
molecules, showing $\eta^5$ coordination;  (c) phthalocyanine molecule.}
\end{figure}

Figs. 1(a) and 1(b) show the eclipsed and staggered geometrical structures of metallocene
molecules, which have the standard $\eta^5$ coordination, whereby the central metal ion bonds
with all five carbon atoms in each ring. There exist other possible 
geometrical arrangements, where the rings are displaced laterally, or even tilted, so that only
one, two or three carbons in each ring bond to the metal ion; the coordination is called 
$\eta^{1,2,3}$, accordingly. All of the metallocenes that we have simulated have been synthesized,
except for AuCp$_2$, and possibly CuCp$_2$. Their geometrical structures for the gas phase have been
determined experimentally for Fe, Co, Ni and Zn\cite{Haaland79,Garkusha98}. Detailed quantum 
chemistry simulations of the geometry and total spin of isolated Fe, Co, Ni, Cu and Zn metallocenes
have been
performed\cite{Xu03,Lissenko03,Liu03,Swart07}, which agree with the experimental data. 
Our results 
both for the geometry and the spin ground state, which are summarized in Table I, agree in detail
with the available experimental and theoretical data. 

\begin{table}
\begin{tabular}{|c||c|c|c|c|}
\hline
Molecule  & Exp. geometry &  Our geometry      & $M_T$ ($\mu_B$) & $M_{M}$ ($\mu_B$) 
\\ \hline \hline
FeCp$_2$ & Eclipsed $\eta^5$	     &  Agree		           & 0.0      & 0.0
\\ \hline
CoCp$_2$ & Eclipsed $\eta^{1,2}$     &  Agree		           & 0.99     & 0.75
\\ \hline
NiCp$_2$ & Eclipsed $\eta^5$	     &  Agree		      	   & 1.99     & 1.22
\\ \hline
CuCp$_2$ & Not synthezised      &  Staggered $\eta^2-\eta^5$  & 0.99     & 0.22
\\ \hline
ZnCp$_2$ & Staggered $\eta^1-\eta^5$ &  Agree		           & 0.0      & 0.0
\\ \hline
IrCp$_2$ & Not available	     &  Eclipsed	           & 0.95     & 0.50
\\ \hline
PtCp$_2$ & Not available	     &  Staggered $\eta^3-\eta^3$  & 0.0      & 0.0
\\ \hline
AuCp$_2$ & Not synthezised      &  Staggered $\eta^2-\eta^2$  & 0.99     & 0.13
\\ \hline
\end{tabular}
\caption{Structural and magnetic properties of the metallocene molecules
simulated in this article. The total spin moment 
of the molecules is denoted by $M_T$ and is oriented along the z-axis. The spin moment of
the transition metal atom $M_M$ is estimated from a Mulliken population analysis. }
\end{table}

We select the cobalt, copper and gold metallocenes for a closer inspection of the magnetic phenomena
of these molecules. The rationale behind our choice is the following. First, the three molecules
have the same spin (S=1/2) because of the axial crystal field of the molecule, so that their magnetic
properties can be meaningfully compared. Second, cobalt is a strong ferromagnet, while copper and
gold have a noble metal electronic shell structure. Third, gold has more delocalized s-wave
functions, so it is expected to show a weaker magnetism. Fourth, we wish to understand better the
magnetic properties of small gold molecules in view of the recent experiments mentioned
above\cite{Crespo04,Yamamoto04,Garitaonandia08}.
We plot in Fig. 2(a) a schematic view of the geometrical structure and atomic spin moments of
CoCp$_2$. We have found that the most stable isomer has basically $D_{5h}$ symmetry, with small
Jahn-Teller distortions, which favor $\eta^2$ coordination. 
More important for the
present discussion is the crude size of the atomic moments in the molecule, which the figure
shows to be localized in the cobalt atom. Table I indeed shows that Co carries 75\%
of the total moment of the molecule ($M_{Co}=0.75 \mu_B$). Notice that the $\eta^1-\eta^2$ 
coordination is reflected in the small moment of two of the carbon atoms in each carbon ring.
One of the two atom rings in the $D_{5d}$ geometry CuCp$_2$ has been displaced and shows $\eta^2$;
the other ring shows a much more slight displacement, which leads to $\eta^3$ coordination, as we
show in Fig. 2(b). Furthermore, the two
rings are not coplanar. The magnetic configuration of this molecule is rather interesting.
The transition metal atom has a much lower magnetic moment, which amounts to only 20 \% of the total 
moment of the molecule, see Table I. Fig 2(b) shows that the magnetization of the carbon atoms 
follows a non-monotonic dependence with the distance to the copper atom, which is reminiscent of
a sort of spin-density wave. Indeed, two of the carbon atoms in the displaced ring have spin moments
of 0.25 $\mu_B$, while the other three are very weakly magnetized ($M\sim 0.02 \mu_B$). A similar
situation happens in the other ring, where two carbon atoms have sizable magnetizations of about
$0.13 \mu_B$, while the moments of the other three are tiny.
This unconventional behavior is more apparent in AuCp$_2$, that we show in Fig. 2(c). The two
cyclopentadyenil rings are displaced in opposite directions by the same distance, which leads to
$\eta^2-\eta^2$ coordination. The two rings are coplanar in this case, which makes a much more
symmetric molecule than CuCp$_2$. The figure clearly indicates that two carbon atoms in each ring
have a much larger magnetization ($0.21 \mu_B$) than the one shown by the central gold
atom ($0.13\mu_B$). The magnetization profile shows a more pronounced non-monotonic behavior than
CuCp$_2$: the two carbon atoms in each ring which are directly bonded to the gold atom are
unmagnetized, while the carbon atom furthest apart shows antiparallel magnetization. This suggest even
more strongly than in the CuCp$_2$ case that we are facing with a spin-density wave profile. 

\begin{figure}
\includegraphics[width=\columnwidth]{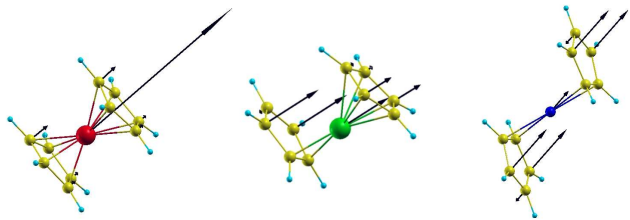}
\caption{(Color online) Geometry and atomic spin moments of the most stable isomers of (a) CoCp$_2$;
(b) CuCp$_2$; and (c) AuCp$_2$. 
Notice that the energy is unaffected by a global spin rotation, since the magnetic anisotropy
energy is essentially zero in the three molecules. We have chosen to display here a case were
the spins are oriented perpendicular to the cyclopentadienyl rings.}
\end{figure}

We note now that we have used a Mulliken population analysis to draw the above conclusions on the 
magnetism of metallocene molecules. To dispel fears on the possible inaccuracies of this analysis,
we have estimated also the atomic charge  and spin moments by integrating the charge and spin
densities inside spheres centered in each atom, for a number of different integration
radii, that we show in Fig. 3(a). We find that the spin density 
curves show plateaus for radii slightly larger than the inter-atomic distances. The values of the
spin moments at these plateaus coincide almost quantitatively with the estimates provided by the
Mulliken analysis. The plateaus are not present for the atomic  charges, apart from a small shoulder 
in the case of gold. To understand better the spatial distribution of the integrands, we plot in Figs.
3(b) and (c) the density iso-contours. The charge iso-contours show the charge conjugation appropriate
to the covalent $\pi$ chemical bond among the $p_z$ orbitals in the carbon rings. This is the reason
why the integrated charges of the carbon atoms do not show plateau behavior. In contrast the charge
about the central gold atom has a spherical shape, which points towards a more ionic character of the
gold-carbon bond. This is also reflected in the shoulder referred above. 
The spin iso-contours have a lobular structure. The four lobes that appear for Au correspond
to the $d_{zy}$ orbital, which is the most directly involved in the gold-carbon chemical bond
responsible for the $\eta^2$ coordination. The lobular structure of the carbon atoms corresponds to
the $p_z$
orbitals; the lobes point slightly towards the adjacent carbon atoms, which indicates
a larger delocalization due to the conjugation present in the ring. We note that the atoms that
showed antiparallel spin moments in each ring in Fig. 2(c), show here a negative spin density.
The localization of the spin density in the orbitals making up the chemical bonds in the molecule,
together with the non-uniform variation of the atomic moments strongly suggest again
that the spin landscape in this molecule corresponds to a static spin density wave originated by the
presence of the gold atom. 

\begin{figure}
\includegraphics[width=\columnwidth]{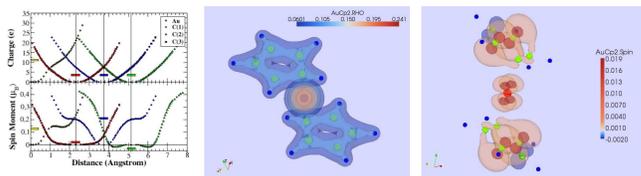}
\caption{Atomic spin moments and charge of AuCp$_2$. (a) Charge and spin moments estimated
by integrating the charge and spin densities on spherical contours. The abscissa in the graph 
corresponds to an imaginary line that begins at the Au atom, and goes to one of the cyclopentadyenil
rings, where it passes first through the closest carbon atom(C(1)), then through the following one
(C(2)) and ends up on the farthest (C(3)). The position of each of these carbon atoms is marked by
a vertical line in the graph. The colored rectangular bars indicate the Mulliken population results
for the charge and spin moments for each atom. The different data points correspond to increasing
integration radii for the spherical integration contours. (b) and (c) Charge and spin iso-contours
of the molecule.}
\end{figure}

\begin{figure}
\includegraphics[width=\columnwidth]{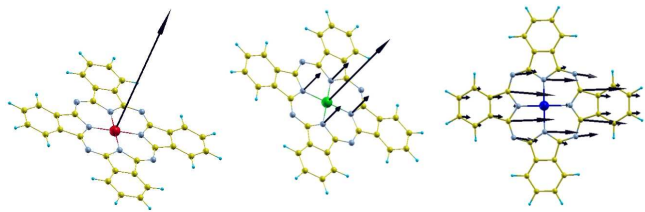}
\caption{(Color online) Geometry and atomic spin moments of the most stable isomers of (a) CoPc; (b)
CuPc; and (c) AuPc.
The plane of the CoPc and CuPc molecules is tilted in the figure on purpose.
Notice that the spins in AuPc show small non-collinearities.}
\end{figure}

We finally note that the magnetic moments are essentially collinear. 
We have also computed the expectation value of the atomic orbital
moments, which we find to be 0.02-0.03 $\mu_B$ at the gold atom, and below 0.001 $\mu_B$ at most for
the surrounding
carbon atoms (as otherwise expected since the spin orbit coupling constant is only large for Au).
We therefore find that the ratio $M_L/M_S$ is of the order 0.2-0.3 at most. This conclusion
agrees with the experimental results for gold nanoparticles\cite{Garitaonandia08,Yamamoto04}.

\begin{figure*}
\includegraphics[width=\textwidth]{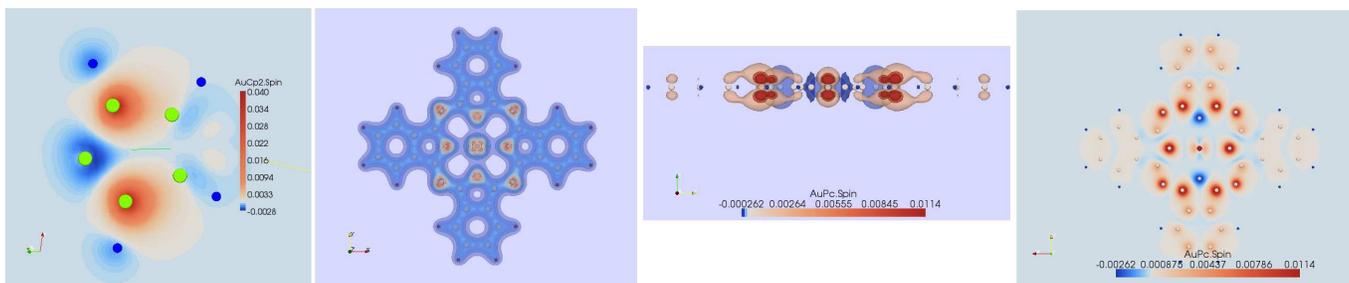}
\caption{
(a) A top view of the spin density wave in one of the two rings of orocene. Red
(blue) colour indicates positive (negative) spin densities. The colour intensity indicates how
positive or negative is the spin density. (b) Top view of the charge iso-contours in a gold
phthalocyanine. (c) and (d) Lateral and top views of the spin density wave in AuPc. The colour code
is the same as in AuCp$_2$.}
\end{figure*}

The above trends on the magnetic behavior of metallocenes molecules
can be generalized to other organic molecules containing transition metal ions. like phthalocyanines.
We have found that the geometry of these molecules does not 
change during the force relaxation cycle, and only the positions of the central atoms are slightly
modified. Cobalt, copper and gold phthalocyanines (CoPc, CuPc and AuPc, respectively) all have spin
1/2 and zero magnetic anisotropy, as we desired. We have 
therefore chosen to plot in Fig. 4 our results for molecules laying in the $XY$ plane, with spins
oriented along the z-axis for CoPc and CuPc; for AuPc in contrast,
we found it more illustrative to plot a case where the spins were oriented in
the plane of the molecule. and along the x-axis for AuPc. We find that the atomic spin moments are
very strongly localized at the cobalt atom for CoPc, but that they are spread towards the
neighboring
nitrogen atoms in CuPc. Gold phthalocyanine is specially interesting again. The spin moments are
spread at the central core and all along the benzene rings in one of the arms of the molecule, 
while the moments of the central gold atom and of the  benzene rings at the other arm are
negligible. Hence, the spin configuration in the ground state breaks the $C_{4v}$ symmetry of
the molecule down to $C_2$. We stress that the moment at the gold atom is negligible, and
that the role of this atom seems to produce a spin-density wave, magnetizing its surrounding atoms.
As in gold metallocene, we find that the orientation of the magnetic moments in AuPc is collinear 
to a first approximation. Interestingly, we have found slight non-collinearities when we orient the
spins in the plane of the molecule, as is apparent in Fig. 4(c). In contrast, the non-collinearities 
dissappear when the orientation of the atomic spins is perpendicular to that plane.

We finally wish to pay a closer look at the spin-density profile of AuCp$_2$ and AuPc. We plot to
this end the spin density in one of the cyclopentadyenil rings of orocene, in Fig. 5(a).
We find that this density is largely spread throughout the ring, as opposed to being strongly
localized about the carbon atoms, and shows wide regions of opposite charge. We plot in Fig. 5(b)
the charge density of AuPc, which shows the covalency of the benzene rings, and the more ionic
character of the bonds linking nitrogen and gold atoms. Fig. 5(c) shows a lateral view
of the spin density in AuPc,
which indicates that the magnetization is localized in the $p_z$ nitrogen and carbon orbitals.
A top view of the molecule in Fig. 5(d) indicates that the spin density is also delocalized in
this molecule, showing small regions of negative spin surrounded by much larger regions where
the spin density is positive.

As a summary, we find that small gold organic molecules show unconventional magnetic features.
We find that the role of the gold atom is to magnetize the
surrounding atoms, while displaying themselves small or even negligible magnetic moments. The
magnetization
pattern shows a spin-density wave behavior, whereby the spin profile is fairly delocalized across the
molecule, and presents regions with positive and with negative spin density. We find that the
orientation of the spins might be parallel or anti-parallel, but is collinear to a first 
approximation. We also find that the orbital moments are much smaller that the spin moments.

We are indebted to I. K. Sch\"uller and F. J. Garc\'{\i}a Alonso. 
We acknowledge financial support from the Spanish Research Office (project FIS2006-12117).

\end{document}